\documentclass[aps,twocolumn,superscriptaddress,longbibliography,numerical]{revtex4-1}

\usepackage{amsmath}
\usepackage{epsfig}
\usepackage{graphicx}
\usepackage{bm}
\usepackage{amssymb}
\usepackage{slashed}
\usepackage{hyperref}
\hypersetup{
     colorlinks   = true,
     citecolor    = blue,
     urlcolor = blue,
     linkcolor = blue
}

\begin{document}


\title{Nonlinear Chiral Transport in Dirac Semimetals}

 \author{Alexander~A.~Zyuzin}
  \affiliation{Department of Applied Physics, Aalto University, P.~O.~Box 15100, FI-00076 AALTO, Finland}
  \affiliation{Ioffe Physical--Technical Institute,~194021 St.~Petersburg, Russia}

 \author{Mihail Silaev}
 \affiliation{Department of Physics and Nanoscience Center, University of Jyv\"askyl\"a,
P.O. Box 35 (YFL), FI-40014 University of Jyv\"askyl\"a, Finland}

 \author{Vladimir~A.~Zyuzin}
 \affiliation{Department of Physics and Astronomy, Texas A$\&$M University, College Station, Texas 77843-4242, USA}


\begin{abstract}
We study the current of chiral charge density in a Dirac semimetal with two Dirac points in momentum space, subjected to an externally applied time dependent electric field and in the presence of a magnetic field. 
Based on the kinetic equation approach, we find contributions to the chiral charge current, that are proportional to the second power of the electric field and to the first and second powers 
of the magnetic field, describing the interplay of the chiral anomaly and the drift motion of electrons moving under the action of electric and magnetic fields. 
\end{abstract}
\maketitle

\section{Introduction}
The Weyl and Dirac semimetals are recently discovered materials, whose conduction and valence bands with linear energy dispersion touch at a number of Weyl or Dirac points in the Brillouin zone \cite{Abrikosov, Volovik_review, Murakami, Savrasov, Second_Weyl, Burkov_Balents}. These systems belong to the Fermi point universality class of fermionic vacua \cite{Volovik_review} and possess nontrivial topology of the electronic band structure. The non degenerate Weyl point might be described as a monopole sink or source of the Berry curvature and assigned with a topological charge, an integral of the Berry curvature over the surface enclosing the point. Since the net topological charge is zero, Weyl points always appear in pairs of opposite charge. The Dirac point might be composed of two Weyl points with topological charges of opposite sign. In certain classes of three-dimensional semimetals such Dirac points occur in pairs separated along a rotation axis of the crystal provided both time-reversal and inversion symmetries are not broken \cite{Morimoto, Nagaosa, Gorbar_Dirac, Furusaki}. 

One of the distinct properties of Weyl and Dirac semimetals is the chiral anomaly, which is a non-conservation of chiral charge induced by the externally applied parallel electric and magnetic fields \cite{Adler, Bell_Jackiw}. The presence of the chiral charge imbalance leads to a number of phenomena such as for example the chiral magnetic effect - charge current driven along the magnetic field \cite{Kharzeev_CME}, chiral electric separation effect - the flow of chiral charge imbalance along the electric field \cite{Kharzeev_review}, the quantum and classical negative magnetoresistance \cite{Nielsen_Ninomiya, Son_Spivak, Burkov_negative_MR, Kharzeev, Vova1, Spivak_Andreev}, and contributions to the nonlinear optical response \cite{SHG, Tanaka_PRB, Morimoto_PRB, Cortijo_PRB, Zyuzin_SHG, Konig_PRB, Juan_PRB, Rostami_PRB}.  
Another anomalous transport phenomena, although unrelated to the chiral anomaly, is the chiral separation effect, which describes the flow of fermions with opposite chiral charges in opposite directions along with the external magnetic field \cite{Zhitnitsky1, Zhitnitsky2}. The progress in the topological semimetals is reviewed in Ref. \cite{Mele_Review}.

Recently, a question of the interplay of the chiral anomaly and the nonlinear chiral transport was addressed for a ferromagnetic Weyl semimetal \cite{Vova2}. 
Based on the kinetic equation approach \cite{Berry_review, Son_Yamomoto}, it was shown that the chiral anomaly might lead to quadratic in electric field corrections to the chiral charge current. 

Here, we study the chiral charge current driven by a time-dependent electric field in the presence of a magnetic field in the Dirac semimetal, with a pair of Dirac points in it's band structure.
Besides the chiral charge imbalance, the chiral anomaly generates a spin imbalance in each Dirac valley, such that the total spin polarization in the system is zero, although the staggered spin polarization is induced.
We show that the chiral charge current as well as the current of staggered spin polarization is proportional to the second power of the electric field and is described by joint action of the chiral anomaly and the electron motion in the presence of the electric and magnetic fields.

\section{Model} Let us consider a model of the inversion and time reversal symmetric gapless Dirac semimetal with two Dirac points separated in momentum space on the crystal rotation axis (one might have in mind $\textrm{Cd}_3\textrm{As}_2$ and $\textrm{Na}_3\textrm{Bi}$ as particular material candidates).
The system is described by the Hamiltonian
\begin{equation}\label{Ham}
H(\mathbf{k}) =v (\sigma_x s_z k_x - \sigma_y k_y) + m(k_z)\sigma_z + \delta H(\mathbf{k}) ,
\end{equation}
where $m(k_z) =m_1 k_z^2 -m_0$, in which $m_0m_1>0$, and $\boldsymbol{\sigma}$ and $\mathbf{s}$ are the vectors composed of the three Pauli matrices denoting the pseudo-spin and spin degrees of freedom (we set $\hbar=1$). 
The Hamiltonian $\delta H(\mathbf{k})  = \gamma \sigma_x k_z (s_{+}k_{-}^2+s_{-}k_{+}^2)\propto \mathcal{O}(k^3)$ is a small correction, which is off-diagonal in spin space. 
Two Dirac points are separated by a distance $2\sqrt{m_0/m_1}$ along $z$-axis in momentum space. Provided $\delta H(\mathbf{k}) =0$ the Hamiltonian in Eq. \ref{Ham} is block diagonal 
and one can introduce a sign, $s=\pm$, to label the eigenvalues of $s_z$. 

To proceed, we consider a spherical Fermi surface, set $2\sqrt{m_0m_1} \equiv v$, and linearize the Hamiltonian around each Dirac point as
$
H_{\eta,s}(\mathbf{k}) = v (s \sigma_x k_x -\sigma_y k_y+\eta \sigma_z k_z ) ,
$
where the momentum in each valley is now measured relatively to the corresponding point, which is labeled by $\eta = \pm$, as $k_z \rightarrow k_z -\eta \sqrt{m_0/m_1}$.
We note that each Dirac point is composed of two Weyl points of opposite topological charge, which are related by the time reversal symmetry and determined by the spin eigenvalues.
The Berry curvature for each of four Weyl points is given by
$
\boldsymbol{\Omega}_{\eta,s} = \eta s  \hat{\mathbf{k}}/2k^2,
$
where $\hat{\mathbf{k}}=\mathbf{k}/k$ is the unit vector in the direction of momentum. 

In the absence of the spin-flip processes, the $s_z$-component of the spin is conserved, allowing one to introduce the topological charge for the spin-up and spin-down electrons
$
C_{\eta,+} - C_{\eta,-},
$ with $ C_{\eta, s}  = \int_S d\mathbf{S}\cdot\boldsymbol{\Omega}_{\eta,s}/2\pi$, where the integral is taken over the surface $S$ enclosing the Weyl node. While the total topological charge $\sum_{\eta} ( C_{\eta,+} + C_{\eta,-} )$ is zero,
the staggered spin charge is finite,
$
\sum_{\eta} \eta  (C_{\eta,+} - C_{\eta,-}) /2=2
$; for a more detailed discussion of $\mathbb{Z}_2$ topological charge in the Dirac semimetals, see Refs. \cite{Morimoto, Nagaosa, Gorbar_Dirac, Furusaki}. 

In the situation where a magnetic field is applied to the semimetal, one naturally expects the chiral separation effect.
Turning on an electric field in addition to the magnetic field gives rise to a chiral anomaly with pronounced nonlinear corrections to the chiral charge current. 
This is in contrast to the chiral electric separation effect studied in Refs. \cite{Zhitnitsky1, Zhitnitsky2, Kharzeev_review, Gorbar_Rudenok}, being linear in powers of electric field.  

\subsection{Kinetic equation}
Having established the model of the Dirac semimetal, let us analyze the chiral charge current within the chiral kinetic equation approach focusing on the zero temperature limit. 
This approach has been described extensively in the literature and here we briefly outline the key points \cite{Berry_review, Son_Yamomoto, Niu_PRL, Niu_PRB, Gorbar_2order}. 

We assume a spatially homogeneous time-dependent electric field 
\begin{equation}\label{EB}
\mathbf{E}(t) =  \mathbf{E}_{0}(\omega) e^{-i\omega t}  + \mathbf{E}_{0}^*(\omega) e^{i\omega t}
\end{equation}
and a magnetic field $\mathbf{B}$ applied to the system (we will comment on the effect of the wave-vector dependence of the electromagnetic field later in the conclusions). 
We consider the case of electron doped semimetal, in which the chemical potential is in the conduction band $\mu>0$, neglect the Zeeman effect of a magnetic field compared to its orbital effect, and focus on the response, which is quadratic in powers of electric field. 

The kinetic equation for the distribution function $f_{\eta, s} ( t, \mathbf{k})$ of the wave-packet with energy $\varepsilon_{\eta,s} = \varepsilon(1 -\frac{e}{c}\mathbf{B}\cdot\boldsymbol{\Omega}_{\eta, s})$, where $\varepsilon=vk$, reads
\begin{eqnarray}\label{kinur}
\frac{\partial f_{\eta,s}}{\partial t} + \dot{\mathbf{k}}\cdot \frac{\partial f_{\eta,s} }{\partial \mathbf{k}} = \mathcal{I}[f_{\eta,s}].
\end{eqnarray} 
The electric and magnetic field dependent higher order corrections to the energy and to the Berry curvature do not change the result and will be neglected \footnote{We neglect quadratic in the fields terms to the energy of the wave-packet, Refs. \cite{Niu_PRL, Niu_PRB, Gorbar_2order}. These terms do not contribute to Eq. \ref{Answer_disorder}, while corrections to the Hall contribution to the chiral current \ref{Hall} arising from these terms 
are small $\tau\mu \ll 1$.}.
The kinetic equation is supplemented by the solutions of equations of motion, which contain contributions from the Berry curvature and orbital magnetic moment
\begin{subequations}\label{EqMotion2}
\begin{eqnarray}
\dot{\mathbf{k}} &=&  eD^{-1}_{\eta,s}\left\{ \mathbf{E}+ \frac{1}{c} [\mathbf{v}_{\eta,s}\times \mathbf{B}] + \frac{e}{c}(\mathbf{E}\cdot\mathbf{B})\mathbf{\Omega}_{\eta,s}\right\},~~~~~\\
\dot{\mathbf{r}} &=& D^{-1}_{\eta,s}\left\{ \mathbf{v}_{\eta,s} + e[\mathbf{E}\times\mathbf{\Omega}_{\eta,s}]
+  \frac{e}{c} (\mathbf{v}_{\eta,s}\cdot\mathbf{\Omega}_{\eta,s})\mathbf{B} \right\},~~~~~
\end{eqnarray}
\end{subequations} 
where $\mathbf{v}_{\eta,s} = \partial \varepsilon_{\eta,s} /\partial \mathbf{k}$ is the wave-packet velocity, $\mathcal{I}[f_{\eta,s}]$ is the collision integral, $D_{\eta,s} = 1+\frac{e}{c}(\mathbf{B}\cdot\mathbf{\Omega}_{\eta, s})$, and $e<0$. 

\subsection{Collision integral}
The chiral charge is not strictly conserved when terms nonlinear in momentum and spin-flip scattering processes are included in the Hamiltonian \cite{Burkov_negative}. 
For the collision integral in Eq. \ref{kinur}, we assume that the inter-valley scattering rate is exponentially suppressed with respect to the intra-valley scattering rate. We then note that the Hamiltonian in Eq. \ref{Ham} is block-diagonal in spin-space and the $z$-component of the particle's spin is a conserved quantity provided $\delta H(\mathbf{k})$ is neglected. Turning on the spin-flip processes, we adopt a model in which the intra-valley spin-flip relaxation time is much longer than the intra-valley spin-conserving relaxation time. 
We also assume the magnetic length $v/\sqrt{\omega_c\mu}$, where $\omega_c = -e v^2 B/c\mu$ is the cyclotron frequency, to be much larger than the correlation radius of the scattering potential. Hence the spin-flip and valley-flip relaxation times can be considered magnetic field independent \cite{Bardarson}. These assumptions allow us to simplify the collision integral and separate the intra-valley spin-conserving contribution 
\begin{eqnarray}
\mathcal{I}[f_{\eta,s}]  = \frac{\langle f_{\eta,s}\rangle - f_{\eta,s}(t,\mathbf{k})}{\tau(\varepsilon)} + \Lambda[f_{\eta,s}] + \mathcal{I}_{\mathrm{in}}[f_{\eta,s}],
\end{eqnarray}
where $ \mathcal{I}_{\mathrm{in}}[f_{\eta,s}]$ describes the energy relaxation processes, the valley-flip and spin-flip elastic scattering processes are described by the functional
\begin{eqnarray}\nonumber
\Lambda[f_{\eta,s}] &\equiv&  \frac{\langle f_{-\eta,s} - f_{\eta,s}\rangle}{\tau_{V}(\varepsilon)} + \frac{\langle f_{-\eta,-s} - f_{\eta,s}\rangle } {\tau'_{V}(\varepsilon)}
\\
&+&\frac{\langle f_{\eta,-s} - f_{\eta,s}\rangle}{\tau'_{0}(\varepsilon)} ,
\end{eqnarray}
in which $\tau_{V}(\varepsilon)$ and $\tau'_{V}(\varepsilon)$ are the inter-valley spin-conserving and spin-flip scattering times, $\tau'_{0}(\varepsilon)$
is the intra-valley spin-flip scattering time, and 
\begin{equation}
1/\tau(\varepsilon) = 1/\tau_{0}(\varepsilon) + 1/\tau_{V}(\varepsilon) + 1/\tau'_{0}(\varepsilon) + 1/\tau'_{V}(\varepsilon)
\end{equation}
is the momentum relaxation rate of a particle with energy $\varepsilon$, in which $\tau_{0}(\varepsilon)$ describes the intra-valley spin-conserving scattering processes. We consider the hierarchy of the elastic scattering times $\tau_0(\varepsilon)< (\tau'_{0}(\varepsilon), \tau_{V}(\varepsilon) ) < \tau'_{V}(\varepsilon)$, where the inter-valley spin-flip length $v\tau'_{V}(\mu)$ is assumed to be smaller than the system size. Different inter-node relaxation processes in the Dirac semimetal were studied in detail in Ref. \cite{Abanin_Weyl}.

The triangle brackets $\langle ... \rangle$ mean integration over the directions of momentum, taking into account the change of the phase space in the presence of the magnetic field  \cite{Son_Yamomoto, Berry_review}, such that
$\langle f_{\eta,s}\rangle \equiv \int \frac{d \boldsymbol{\Theta}}{4\pi} D_{\eta,s}(\mathbf{k}) f_{\eta,s}(t,\mathbf{k})$.
The term describing the valley and spin flip elastic scatterings $\Lambda[f_{\eta,s}]$ satisfies $\sum_{\eta,s} \Lambda[f_{\eta,s}] = 0 $ and can be related to the distribution function as  
$\sum_{\eta, s} \eta s \left\{\Lambda\left[f_{\eta,s}\right] +\langle f_{\eta,s}\rangle/\tau_f(\varepsilon) \right\} = 0$,
where 
$
1/2\tau_f(\varepsilon) \equiv 1/\tau_{V}(\varepsilon) + 1/\tau'_{0}(\varepsilon) 
$
is an effective relaxation rate, which includes inter-valley spin-conserving and intra-valley spin-flip scattering processes. 

\section{Solution of kinetic equation}
Let us now study the chiral charge response of the Dirac semimetal taking into account the inter and intra valley scattering processes. To reveal the topologically nontrivial contributions 
we consider the limit of weak magnetic field, in which the cyclotron frequency is smaller than the electron momentum relaxation rate. 

We search for the approximate solution of Eq. \ref{kinur}, keeping contributions to the distribution function up to second order of the electric field. 
Similarly to Ref. \cite{Glazov}, we the expand distribution function in powers of the incident electric field,
\begin{align}\label{DF_expansion}\nonumber
f_{\eta,s}(t,\mathbf{k}) &= f^{(0)}_{\eta,s}(\varepsilon) + \frac{1}{2}\left[ f_{\eta,s}^{(1)}(\omega ,\mathbf{k})e^{-i\omega t} + \textrm{c.c.} \right]\\
&+ \tilde{f}_{\eta,s}^{(2)}(\mathbf{k}) +\frac{1}{2} \left[f_{\eta,s}^{(2)}(2\omega ,\mathbf{k})e^{-2i\omega t} + \textrm{c.c.}\right],
\end{align} 
where $f^{(0)}_{\eta,s}(\varepsilon) \approx  f^{(0)}(\varepsilon) - \frac{e}{c}(\mathbf{B}\cdot\boldsymbol{\Omega}_{\eta, s}) \varepsilon \partial_{\varepsilon} f^{(0)}(\varepsilon)$
and $f^{(0)}(\varepsilon) = \theta(\mu - \varepsilon )$ is the distribution function at zero temperature, $f_{\eta,s}^{(1)}(\omega ,\mathbf{k})$ is the first order correction, and $\tilde{f}_{\eta,s}^{(2)}(\mathbf{k}), f_{\eta,s}^{(2)}(2\omega,\mathbf{k})$ are the second order corrections at the zeroth and double frequencies, respectively. Following Perel' and Pinskii \cite{Perel_Pinski}, we set an additional constrain on the second order correction 
\begin{equation}\label{main_condition}
\sum_{\eta,s}\int d^3k D_{\eta,s} f_{\eta,s}^{(2)}=0, 
\end{equation}
meaning that the second order solution does not change the concentration of particles.

The kinetic equation for the correction to the distribution function in $n$-th ($n>0$) power of the electric field is given by
\begin{align}\label{First_order}\nonumber
&\left(i n \omega - \frac{e }{c}D_{\eta,s}^{-1} [\mathbf{v}_{\eta,s}\times \mathbf{B}]\cdot \frac{\partial}{\partial \mathbf{k}} \right)f_{\eta,s}^{(n)} 
+\mathcal{I}[f^{(n)}_{\eta,s}]
\\
&
=
e D_{\eta,s}^{-1}\left\{\mathbf{E}_{0} + \frac{e}{c}(\mathbf{E}_{0}\cdot\mathbf{B})\mathbf{\Omega}_{\eta,s}\right\} \cdot \frac{\partial  f_{\eta,s}^{(n-1)}}{\partial \mathbf{k}}.~~~~
\end{align}

Physically, the solution to the equation linear in electric field [with $n=1$ in Eq. \ref{First_order}] describes the elastic scattering in the system, while the nonlinear solution should also account for the energy relaxation.
Hence, neglecting the inelastic scattering in the collision integral, the solution of the first order differential equation \ref{First_order} for $f_{\eta,s}^{(1)}(\omega,\mathbf{k})$ is given by
\begin{widetext}
\begin{eqnarray}\label{Solution_1}
f_{\eta,s}^{(1)}(\omega,\mathbf{k}) = \frac{\tau(\varepsilon)}{1-i\omega\tau(\varepsilon)} 
\bigg[
\frac{\langle f_{\eta,s}^{(1)}\rangle}{\tau(\varepsilon)}+ \Lambda[f_{\eta,s}^{(1)}] 
- 
\frac{ev}{D_{\eta,s}} 
\bigg\{ \frac{\eta s e}{2 ck^2} (\mathbf{E}_{0} \cdot \mathbf{B})+
\mathbf{E}_{0}\cdot \hat{\mathbf{k}} 
- 
\kappa_{\eta,s} \frac{ [\mathbf{E}_{0} \times  \hat{\mathbf{B}}]
+  \kappa_{\eta,s} \mathbf{E}_{0} }{1 +\kappa^2_{\eta,s}} \cdot \hat{\mathbf{k}}_{\perp}
\bigg\} 
\partial_{\varepsilon} f^{(0)}_{\eta,s} 
\bigg],~~~~~
\end{eqnarray}
\end{widetext}
where parameter $\kappa_{\eta,s}(\varepsilon,\omega) = \frac{\mu}{\varepsilon}\omega_cD_{\eta,s} \frac{\tau(\varepsilon)}{1-i\omega\tau(\varepsilon)}$ is introduced for brevity, $\hat{\mathbf{k}}_\perp$ is the unit vector in the direction of momentum lying in the plane transverse to the direction of magnetic field $\hat{\mathbf{B}}=\mathbf{B}/B$. 
The last term in Eq. \ref{Solution_1} absorbs the contribution from the cyclotron part of the Lorentz force. 
It is worth to note that the cyclotron frequency in $\kappa_{\eta,s}(\varepsilon,\omega)$ is renormalized with the Berry curvature. 

Multiplying Eq. \ref{Solution_1} with $D_{\eta,s}$ and integrating over the directions of momentum, one obtains an equation for $\langle f_{\eta,s}^{(1)}\rangle$ in the form
\begin{eqnarray}\label{relation_1b}\nonumber
-i\omega \langle f_{\eta,s}^{(1)}\rangle &=&  \Lambda[f_{\eta,s}^{(1)}] - \eta s \frac{e^2v}{2 ck^2} (\mathbf{E}_{0}\cdot\mathbf{B})\partial_{\varepsilon} f^{(0)}\\
&+& \eta s\frac{e^2v}{6ck^2}(\mathbf{E}_{0}\cdot\mathbf{B})\left[\partial_{\varepsilon} f^{(0)}+\varepsilon \partial^2_{\varepsilon} f^{(0)} \right],~~
\end{eqnarray}
where the second line describes the contribution of the orbital magnetic moment.
We then find that the relaxation of the chiral imbalance
\begin{equation}\label{relation_2}
\sum_{\eta, s}\eta s \langle f_{\eta,s}^{(1)}\rangle = - \frac{2 e^2v}{3 ck^2}\frac{ \tau_f(\varepsilon) (\mathbf{E}_{0}\cdot\mathbf{B}) }{1-i\omega\tau_f(\varepsilon)}
[ 2\partial_{\varepsilon} f^{(0)} - \varepsilon \partial^2_{\varepsilon} f^{(0)} ],
\end{equation}
comes from the inter-valley spin-conserving and intra-valley spin-flip scattering processes and describes the emergence of non-equilibrium chiral charge as well as staggered spin accumulations. 
This quantity is associated with the chiral anomaly. 
The staggered spin polarization in the Dirac semimetal, 
$ 
\mathcal{P}(t) = N^{-1}\sum_{\eta,s}\eta s \int (dk)D_{\eta,s} f_{\eta,s}(t,\mathbf{k}),
$ 
where $N=2\mu^3/3\pi^2v^3$ is the electron density, in the lowest order in the electric and magnetic fields
$\mathcal{P}(\omega) = \frac{3\omega_c}{2\mu} \frac{ev\tau_f}{\mu} \frac{(\mathbf{E}_{0} \cdot\hat{\mathbf{B}})}{1-i\omega\tau_f}$ is determined by the $(\mathbf{E}_{0}\cdot \mathbf{B})$ product.
Finally, the sum  $\sum_{\eta, s}  f_{\eta,s}^{(1)}$ gives the standard expression for the first-order solution, which satisfies $\sum_{\eta, s} \langle f_{\eta,s}^{(1)} \rangle$ =0.

The exact solution in the second order is rather cumbersome. However, we are interested in the limit of weak magnetic field $\omega_c < \omega $ and keep up to quadratic in powers of $\omega_c\tau$ corrections to the distribution function. The solution can be formally written as
\begin{align}\label{Solution_2}
f_{\eta,s}^{(2)}(2\omega,\mathbf{k}) &= \frac{\tau(\varepsilon)}{1-2i\omega\tau(\varepsilon)}\bigg\{ 
\frac{\langle f_{\eta,s}^{(2)}\rangle}{\tau(\varepsilon) }  +\mathcal{I}_{\mathrm{in}}[f^{(2)}_{\eta,s}]
\\\nonumber
&-
eD_{\eta,s}^{-1}
\left(1-\frac{e}{c }\tau(\varepsilon)D_{\eta,s} \frac{v[\hat{\mathbf{k}}\times \mathbf{B}]}{1-2i\omega\tau(\varepsilon)}\cdot \frac{\partial}{\partial \mathbf{k} }\right) 
\\\nonumber
&\times
\left[\mathbf{E}_{0} + \frac{e}{c}(\mathbf{E}_{0} \cdot \mathbf{B})\mathbf{\Omega}_{\eta,s}\right]\cdot \frac{\partial}{\partial \mathbf{k} } f_{\eta,s}^{(1)}(\omega,\mathbf{k})
\bigg\},~~
\end{align}
where we neglect $\Lambda[f_{\eta,s}^{(2)}]$, since $\sum_{\eta, s}\eta s \langle f_{\eta,s}^{(2)}\rangle \propto \frac{\omega_c^2}{\mu^2}(\mathbf{E}_0\cdot \hat{\mathbf{B}})^2 \ll 1$ is beyond the validity of our assumptions.
The difference between the momentum relaxation rates of the first and second harmonics is also neglected. 

To determine the contribution to $\langle f_{\eta,s}^{(2)}(2\omega)\rangle \propto \mathbf{E}_0^2$ one has to take into account the inelastic processes \cite{Perel_Pinski} and apply a condition of a constant concentration of particles in the presence of the electric field. For the collision integral, describing the inelastic processes, we consider the simplest form
\begin{equation}
I_{\mathrm{in}}[f^{(2)}_{\eta, s}] = -\frac{f^{(2)}_{\eta, s}}{\tau_{\mathrm{in}}}, 
\end{equation}
where $\tau_{\mathrm{in}}$ is the energy relaxation time of the electron due to coupling to some thermal bath. Taking into account only the terms that give dominant contribution to the chiral charge current, we find
\begin{eqnarray}\nonumber
\langle f_{\eta,s}^{(2)}(2\omega)\rangle &=& \frac{e^2v^2}{3}\frac{\tau_{\mathrm{in}}(\mathbf{E}_0\cdot \mathbf{E}_0)}{1-2i\omega \tau_{\mathrm{in}}} \bigg[\partial_{\varepsilon}\frac{\tau(\varepsilon)}{1-i\omega\tau(\varepsilon)}\partial_{\varepsilon} f^{(0)} \\
&+&\frac{\tau(\varepsilon)}{1-i\omega\tau(\varepsilon)} (\partial_{\varepsilon}^2f^{(0)} + 2\varepsilon^{-1}\partial_{\varepsilon} f^{(0)} )\bigg].
\end{eqnarray} 
We then obtain the second order correction in the form
\begin{eqnarray}\label{Solution_2}\nonumber
f_{\eta,s}^{(2)}(2\omega,\mathbf{k}) &=& 
\langle f_{\eta,s}^{(2)}(2\omega)\rangle
-\frac{ev\tau(\varepsilon)}{1-2i\omega\tau(\varepsilon)}
\hat{\mathbf{k}}\cdot   \bigg
\{
\mathbf{E}_{0}
\\
&+&
\frac{ev}{ck}\frac{\tau(\varepsilon) [\mathbf{E}_{0}\times \mathbf{B}] }{1-2i\omega\tau(\varepsilon)}
\bigg
\} 
\partial_\varepsilon \langle f_{\eta,s}^{(1)}(\omega,\mathbf{k})\rangle,~~
\end{eqnarray}
where we substitute $f_{\eta,s}^{(1)}(\omega, \mathbf{k})$ with a solution of Eq. \ref{relation_1b}. The expression for $\tilde{f}_{\eta,s}^{(2)}(\mathbf{k})$ can be found from Eq. \ref{Solution_2} with the formal substitutions $\omega =0$, $\mathbf{E}\cdot \mathbf{E} \rightarrow |\mathbf{E}|^2$, and $\mathbf{E}_{0}  \langle f_{\eta,s}^{(1)}(\omega,\mathbf{k})\rangle \rightarrow \mathrm{Re} \mathbf{E}_{0}^*  \langle f_{\eta,s}^{(1)}(\omega,\mathbf{k})\rangle$. We are now in the position to calculate the chiral charge current density.

\section{Nonlinear chiral charge current} 
Let us define the chiral charge current density in the semimetal 
$
\mathbf{j}_5(t) = e\sum_{\eta,s} \eta s  \int  \frac{d^3k}{(2\pi)^3} D_{\eta,s} \dot{\mathbf{r}}(t,\mathbf{k}) f_{\eta,s}(t,\mathbf{k})
$, where explicitly 
\begin{eqnarray}\label{Current_def}\nonumber
\mathbf{j}_5(t) &=& e\sum_{\eta,s}  \int  \frac{d^3k}{(2\pi)^3} \bigg\{ \eta s v\hat{\mathbf{k}} + \frac{ev\mathbf{B}}{2ck^2} +  \frac{e}{2k^2}[\mathbf{E}(t)\times\hat{\mathbf{k}}]
\\
&+& \frac{ev}{ck^2}\hat{\mathbf{k}}(\mathbf{B}\cdot \hat{\mathbf{k}})- \frac{ev\mathbf{B}}{2ck^2} 
 \bigg\} f_{\eta,s}(t,\mathbf{k}).
\end{eqnarray}
The first term gives a finite
contribution provided the staggered spin polarization is induced.
The second and third terms are the corrections to the motion of the wave-packet due to the nontrivial Berry curvature and describe the chiral separation and inverse Faraday effects, respectively.
The terms on the second line describe corrections from the orbital magnetic moment of the wave-packet.

We first consider the chiral transport in the collisionless limit and neglect $\mathcal{I}[f_{\eta,s}]$ in Eq. \ref{kinur}.  
The chiral charge current density can be expanded in powers of the electric field amplitude
\begin{eqnarray}\label{Current_expansion}
\mathbf{j}_5(t) = \frac{e^2\mu}{\pi^2 c} \mathbf{B} + \left[  \mathbf{j}_5(2\omega)e^{-2i\omega t} + \textrm{c.c.} \right]/2,
\end{eqnarray}
where the first term describes the chiral separation effect (for a review see Ref. \cite{Kharzeev_review}).
It is a non-dissipative chiral current, which exists in the equilibrium state of chiral fermions provided a magnetic field is applied \cite{Zhitnitsky1, Zhitnitsky2}, and doesn't depend on the orbital magnetic moment.
The other terms describe second-order correction at the double frequency of the electric field. 

In the collisionless limit at frequencies $\omega \gg \omega_c$, we obtain
\begin{align} \label{Answer_clean}
\mathbf{j}_{5}(2\omega) = - \frac{e^3 \omega_c}{6\pi^2 \omega^2} [\mathbf{E}_0\times[\hat{\mathbf{B}}\times\mathbf{E}_0]].
\end{align}
We find that $\mathbf{j}_{5}(2\omega)$ vanishes for the parallel orientation of electric and magnetic fields (the field dependence in Eq. \ref{Answer_clean} matches the one derived for the strain induced non-equilibrium spin current in the Dirac semimetal \cite{Araki}) and does not depend on the orbital magnetic moment. Although, turing on the scattering processes, different inter-valley and intra-valley relaxation rates give rise to a finite chiral current for parallel orientation of the fields.

At $\omega\tau < 1$, where $\tau \equiv \tau(\mu)$ is the momentum relaxation rate at the Fermi energy, we find a linear in magnetic field contribution to the chiral charge current density in the form
\begin{eqnarray} \label{Answer_disorder}\nonumber
\mathbf{j}_{5}(2\omega)&=& 
\frac{2e^3}{3\pi^2}  \frac{\omega_c\tau}{1-i\omega\tau}
\bigg\{
\frac{\tau_{\mathrm{in}}}{1-2i\omega \tau_{\mathrm{in}}} (\mathbf{E}_0\cdot \mathbf{E}_0) \hat{\mathbf{B}}\\
&-& \frac{G(\omega)}{1-2i\omega\tau}\frac{\tau_f}{1-i\omega\tau_f}\mathbf{E}_0 (\mathbf{E}_0 \cdot\hat{\mathbf{B}}) 
\bigg\},
\end{eqnarray}
where the model dependent coefficient $G(\omega)$ is determined by the scattering potential
\begin{equation}
G(\omega) = (1-2i\omega\tau)\frac{1}{2\mu\tau}\partial_\varepsilon\frac{\varepsilon^2\tau(\varepsilon)}{1-2i\omega \tau(\varepsilon)}\bigg|_{\varepsilon = \mu}.
\end{equation}
In the model of intra-valley short range potential, where $\tau_0(\varepsilon) \propto \varepsilon^{-2}$, assuming other scattering times to be energy independent, one estimates $G(0) \propto 1-\tau/\tau_{0} $. Although, for the 
model of the Coulomb impurities $\tau_0(\varepsilon) \propto \varepsilon^2$ one gets $G(0) = 2$. 

The first term in Eq. \ref{Answer_disorder} describes the \emph{ac}-field $\propto \mathbf{E}_0^2$ induced change of the electron
distribution function, being sensitive to the inelastic relaxation time.
Similarly to the chiral separation effect, this contribution to the chiral current is parallel to the magnetic field. 
The second term in Eq. \ref{Answer_disorder} describes the nonlinear chiral electric separation effect, being more pronounced in the case of collinear electric and magnetic fields. 
It is smaller than the first term at $\omega \rightarrow 0$, provided $\tau_{\mathrm{in}} \gg \tau_{f}$. It is also worth noting that the signs of two terms in Eq. \ref{Answer_disorder} 
are opposite. This describes the suppression of the chiral current in the case of collinear fields being stronger at frequencies much larger than the relaxation rates. 

At $\omega\tau<1$ we also find a correction to the chiral current from the interplay of the chiral anomaly and the Hall effect
\begin{eqnarray}\label{Hall}
\mathbf{j}_{5,\mathrm{Hall}}(2\omega) \propto  e^3 \frac{(\omega_c\tau)^2}{1-i\omega\tau} \frac{\tau_f  (\mathbf{E}_0\cdot\hat{\mathbf{B}}) }{1-i\omega\tau_f}  \frac{[\hat{\mathbf{B}}\times\mathbf{E}_0]}{(1-2i\omega\tau)^2}.~~~~~
\end{eqnarray}
This correction is smaller than the drift contribution given in Eq. \ref{Answer_disorder} provided $\omega_c\tau < 1$. 

The physical meaning of the contributions to $\mathbf{j}_5(2\omega)$ can be understood 
if we consider a two-step process. The first step contains long inter-valley scattering, which equilibrates the spin imbalance generated due to chiral anomaly $\propto (\mathbf{E}_0\cdot\mathbf{B})$.
The second step is the drift motion of particles within the valleys under the joint action of the electric and magnetic fields, described by the Lorentz force $e\mathbf{E}+ \frac{e}{c} [\mathbf{v}\times \mathbf{B}] $. This is in contrast to the chiral anomaly generated charge current, which is determined by the inter-valley scattering processes \cite{Son_Spivak, Burkov_negative_MR}. 

The contributions to the axial current in chiral plasma produced by time-dependent electric and magnetic fields and gradients of the chemical potentials were also considered in the Ref. \cite{Gorbar_Rudenok}. While the first term in Eq. \ref{Answer_disorder} is new, the second term and the Hall contribution given in Eq. \ref{Hall} are compatible with the results of Ref. \cite{Gorbar_Rudenok}, provided the chiral chemical potential imbalance is field induced and proportional to $\tau_f (\mathbf{E}\cdot \mathbf{B})$.

\section{Discussion and Conclusions}
So far, we have studied the effect of spatially homogeneous electric field. Briefly, we would like to comment on the case when the electromagnetic wave has a spatial dependence
$\mathbf{E}(t,\mathbf{r}) = \mathbf{E}_0\exp(-i\omega t+ i\mathbf{q}\cdot \mathbf{r}) + \mathrm{c.c.}$
In this case, a finite, linear in electric field contribution to the chiral charge current is allowed. In order to evaluate it, the spatial derivatives in the chiral kinetic equation have to be taken into account, see, for example, Ref. \cite{Gorbar_2order}. Indeed, at $\omega=0$ and using $\tau\ll \tau_f$, which determines the situation where the chiral anomaly is the dominant source to the chiral current, we find that the linear response contribution is given by
$
\mathbf{j}_5(\omega,\mathbf{q}) \propto -ie^2\omega_c \tau \mu\tau_f \mathbf{q}(\mathbf{E}_0\cdot \hat{\mathbf{B}})$. Naturally, it is proportional to the wave-vector $\mathbf{q}$ of the field, such that the chiral current remains invariant with respect to a spatial inversion.

Let us now discuss the experimental feasibility of the proposed effect. The amplitude of the first term in Eq. \ref{Answer_disorder} can be written at zero frequency as $|\mathbf{j}_5| = -\frac{16 e}{3\pi} \frac{\alpha}{\hbar} I \omega_c \tau \tau_{\mathrm{in}}$, where $I = \frac{c\epsilon_0}{2}|\mathbf{E}_0^2|$, $\alpha = e^2/4\pi \epsilon_0 c \hbar $ is the fine structure constant, and $\epsilon_0$ is the permittivity of free space (we have restored $\hbar$ here).
Taking numerical values consistent with experiment Ref. \cite{Ong_experiment}, $\mu\sim 220~\mathrm{meV}$,  $v \sim 9.3\times 10^7~\mathrm{cm/s}$, and $\tau \sim 0.5~\mathrm{ps}$, we estimate $\omega_c \sim 1 ~\mathrm{ps}^{-1}$ at $B=0.2\mathrm{T}$, which satisfies $\omega_c\tau<1$. Although the values of the inelastic relaxation time is not known, we might roughly estimate it as $\tau_{\mathrm{in}} \sim 10 \tau$. We obtain $|\mathbf{j}_5| \sim  50 I [\frac{\mathrm{A/cm^2}}{\mathrm{W/cm^2}}]$, which might be of the order of the quantized circular photogalvanic effect in a Weyl
semimetal studied in Ref. \cite{Juan_PRB}.

The chiral charge current in a Dirac semimetal could be probed indirectly via the interplay between the electric and chiral charge currents, which gives rise to the chiral magnetic waves \cite{Gorbar_2order}.
Although in ferromagnetic Weyl semimetals, where the fields induce a finite spin polarization, probing the chiral charge current might be straightforward via nonlocal measurements similar to the spin polarization in metals \cite{Silsbee, Abanin} (a similar idea proposed for the Weyl and Dirac semimetals was discussed in Ref. \cite{Abanin_Weyl}). However, one needs to be able to extract it from the total signal, which also contains large but electric field and 
frequency independent contribution, described by the chiral separation effect. 

To conclude, we have calculated the nonlinear in the electric field corrections to the chiral charge current in the Dirac semimetal. These are proportional to the second power of the externally applied electric field and consist of contributions that are proportional to the first and second powers of the magnetic field. We have also commented on the chiral anomaly generated staggered spin accumulation, i.e. the nonequilibrium spin polarization in each Dirac valley of the semimetal, with vanishing net spin polarization. 

\begin{acknowledgments}
A.Z. and M.S. are supported by the Academy of Finland.
\end{acknowledgments}
\bibliography{AFM_Weyl_refs_arxiv_resub.bib}

\end{document}